\documentclass[twoside,11pt]{confnl} %%% Please do not change this class !!! %%%
\usepackage{fancyhdr}
\usepackage{amsfonts,amscd,amssymb,amsmath,amsthm,latexsym}
\begin{document}

%% Please modify the following line to include the title of your contribution and acknowledgments:
%Nonlinear and Nonlocal Dynamics of $p$-Adic and Zeta Strings

\title{From $p$-Adic to Zeta Strings\hspace{.25mm}\thanks{\,This article is based on a talk given at the first {\it Conference on Nonlinearity}, held in Belgrade, 11-12 October 2019, and dedicated to the 110th anniversary of Nikolay Nikolayevich Bogolyubov -- an outstanding Soviet and Russian scientist.}}

%% Please modify the following lines to include author names, affiliations and e-mail addresses:

\author{
\bf{Branko Dragovich}\hspace{.25mm}\thanks{\,e-mail address: dragovich@ipb.ac.rs} \\
\normalsize{Institute of Physics Belgrade, University of Belgrade, Belgrade, Serbia}\vspace{1mm} \\
\normalsize{Mathematical Institute, Serbian Academy of Sciences and Arts, Belgrade, Serbia}}
%\vspace{2mm} \\
%\bf{Second author}\hspace{.25mm}\thanks{\,e-mail address: second.author@email.address} \\
%\normalsize{Affiliation of the second author} \vspace{2mm} \\
%\bf{Third author}\hspace{.25mm}\thanks{\,e-mail address: third.author@email.address} \\
%\normalsize{Affiliation of the third author}}

%\date{} %% Please do not modify

\maketitle %% Please do not modify

\begin{abstract}
This article is related to construction of zeta strings from $p$-adic ones.  In addition to investigation of $p$-adic string for a particular prime number $p$, it is also interesting to study  collective effects taking into account all primes $p$. An idea behind this approach is that a zeta string is a whole thing with infinitely many faces which we see as $p$-adic strings. The name zeta string has origin in the Riemann zeta function contained in related Lagrangian. The starting
point in construction a zeta string is Lagrangian for a $p$-adic open  string. There are two types of approaches to get a Lagrangian for zeta string from Lagrangian for $p$-adic strings: additive and multiplicative approaches, that  are related to two forms of the definition of the Riemann zeta function. As a result of differences in approaches, one obtains several different Lagrangians for zeta strings. We briefly discuss  some properties of these Lagrangians, related potentials, equations of motion,  mass spectrum and possible connection with ordinary strings. This is a review of published papers  with some new views.
\end{abstract}

\section{Introduction}

The subject of this paper belongs to String Theory (ST) \cite{GSW}, in particular to $p$-adic and zeta strings. String theory emerged
at the end of 1960th and has been developed as the best candidate for unification of fundamental interactions (gravitational, electromagnetic, strong and weak) and elementary matter constituents in the form of strings. Strings are one-dimensional objects that exist at the very short distances (close to the natural Planck length $\ell_P = \sqrt{\frac{G \hbar}{c^3}} \sim 10^{-35 }\text{m}$). Although strings are not yet experimentally discovered, string theory has played very important role in an interplay between general physical laws and modern mathematics \cite{GSW}.

$p$-Adic strings \cite{freund,VVZ} are introduced in string theory in 1987 as a $p$-adic analog of ordinary strings. In the case of open and closed strings, it was shown that there is a connection between ordinary and $p$-adic strings in the form of product of their scattering amplitudes which is a constant. By construction of Lagrangian for $p$-adic  strings it occurred that $p$-adic strings are even simpler for mathematical investigation than ordinary strings. Interesting properties of $p$-adic strings motivated construction of some other physical models using methods of $p$-adic analysis, and it has resulted  in emergence of {\it $p$-Adic Mathematical Physics}, e.g., see reviews \cite{freund,VVZ,BD1} and references therein.

Note that in the word  ``$p$-adic'', $p$ is related to a prime number. Since there are infinitely many prime numbers, it should be infinitely many  $p$-adic strings. Then question arises how to connect $p$-adic counterparts with ordinary model over real (or complex numbers).  Usual way to connect ordinary model with $p$-adic analogs of the same physical system is by using adelic approach, which is based on adelic analysis. To have insight into adelic aspects of  strings, one can consume \cite{freund,VVZ}. An adelic model of the quantum harmonic oscillator is given in \cite{BD2}.

It seems unnatural that there exist infinitely many kinds of $p$-adic strings -- one string for each prime number $p$.
It is more natural to think that there exists one new entity (say zeta string) which has infinitely many faces -- one face for each $p$. Fortunately, in the case of $p$-adic open  strings there is possibility to work in this direction.  Namely, there are Lagrangians for $p$-adic open strings, which have the same form and the difference is practically in number $p$ that serves as a parameter.
As it will be shown in the sequel of this paper, one can start with this Lagrangian of $p$-adic string, extend $p$ to $n$ (natural numbers), and then by ``integration'' over $n$ to get a new Lagrangian without $n$. This new Lagrangian is related to a new thing, which we call
{\em zeta string}, since this Lagrangian contains the Riemann  zeta function. It occurs that obtaining of this new Largangian
is not unique and depends on applied procedure. This review article is devoted to   construction of  Lagrangians for zeta strings.

The paper is designed as follows. Some basic concepts from $p$-adic mathematics will be
recalled in the next section. A brief review of  $p$-adic open  string, emphasizing related Lagrangian, will be given in section 3. Construction of Lagrangians for zeta strings, and some their elaboration,  is presented in section 4. Several concluding remarks are given in section 5.

This review is  based on a series of papers \cite{BD3}--\cite{BD9}, some of them  published  in the journal {\it Theoretical and Mathematical Physics}, which was founded in 1969 by N. N. Bogolyubov.

\section{A Brief Review of $p$-Adic Mathematical Background}

Recall that rational numbers play important role in physics and mathematics. In physics, numerical results of experiments are rational numbers. In mathematics, rational numbers make a simple infinite number field $\mathbb{Q}$ with respect to summation and multiplication. On $\mathbb{Q}$, in addition to the usual absolute value, there is also well defined $p$-adic norm (or in other words -- $p$-adic absolute value). For a given prime number $p$,  any nonzero rational number can be presented  as $x  = \frac{a}{b} p^\mu$, where $a, b\neq 0$ are integers not devisable  by $p$, and $\mu \in \mathbb{Z}$. Then,
according to definition, $p$-adic norm of $x$ is $|x|_p = p^{-\mu}$ and $|0|_p = 0.$ $p$-Adic distance between two rational numbers $x$ and $y$ is defined as $d_p (x, y) = |x-y|_p .$ $p$-Adic distance is well known example of ultrametrics, because  it satisfies strong triangle inequality, i.e. $d_p(x,y) \leq \text{max} \{d_p(x,z), d_p (y,z)\} .$

Rational numbers, and their completion, with respect to the $p$-adic norm are $p$-adic numbers. Any nonzero $p$-adic number $x$ can be presented in a unique way as
\begin{align}
x = p^\nu \sum_{n=0}^{+\infty} x_n \ p^n , \quad x_0 \neq 0 , \quad \nu \in \mathbb{Z} , \quad x_n \in \{0, 1, 2, \ldots , p-1\} , \label{1.1}
\end{align}
where $x_n$ are digits. For a given $p$, all  numbers \eqref{1.1} make field $\mathbb{Q}_p$ of $p$-adic numbers. There are infinitely many different $\mathbb{Q}_p$ -- one number field for each prime number $p$.

There are mainly two kinds of  analysis with $p$-adic arguments: 1) $p$-adic valued functions  and 2) complex (or real) valued functions. On $\mathbb{Q}_p$  two important continuous complex-valued functions are defined \cite{VVZ}:
\begin{itemize}
\item
1) multiplicative character $\pi_p (x) =
|x|_p^c$, where $x \in \mathbb{Q}_p^\ast = \mathbb{Q}_p \setminus \{0\}$ and $c \in \mathbb{C}$;
\item
2) additive character $\chi_p (x) = e^{2\pi i \{ x \}_p }$, where $x \in \mathbb{Q}_p$ and
$\{ x \}_p$ is fractional part of $x$.
\end{itemize}

Real and $p$-adic numbers have their origin in rational numbers. $\mathbb{Q}$ is dense subset in $\mathbb{R}$ and all $\mathbb{Q}_p$. This fact enables a unification of real and all $p$-adic numbers as a ring of adeles. An adele $(\alpha)$ is an infinite sequence that takes into account real and all $p$-adic numbers:
\begin{align}
\alpha = (\alpha_\infty, \ \alpha_2 , \ \alpha_3 , \ldots , \alpha_p , \ldots ) , \label{2.1}
\end{align}
where $\alpha_\infty \in \mathbb{R}$ and $\alpha_p \in \mathbb{Q}_p , \, \, p =2, 3, 5, 7, \ldots$ with some restrictions.

The following product formulas connect real and $p$-adic properties of the same rational number:
\begin{align}
&|x| \prod_p |x|_p = 1 , \quad \text{if} \, \, x \in \mathbb{Q}\setminus \{0 \} , \\
&e^{-2\pi i x}\prod_p e^{2\pi i \{ x \}_p } = 1 , \quad \text{if} \, \, x \in \mathbb{Q} .
\end{align}
As a comprehensive review on $p$-adic numbers, adeles and their analysis, e.g., see \cite{freund,VVZ,gelfand}.

\section{$p$-Adic Open Scalar Strings}

Recall that ordinary string theory started by construction of the Veneziono amplitude for scattering of two open bosonic strings
$A_\infty (a, b)$, which in the crossing symmetric form is
\begin{align}
 A_\infty (a, b) &= g_\infty^2 \,\int_{\mathbb{R}}
|x|_\infty^{a-1}\, |1 -x|_\infty^{b-1}\, d_\infty x \label{3.1a} \\  &= g_\infty^2 \, \frac{\zeta (1 - a)}{\zeta (a)}\,
\frac{\zeta (1 - b)}{\zeta (b)}\, \frac{\zeta (1 - c)}{\zeta (c)}  ,    \label{3.1b}
\end{align}
where $a, b, c$ are complex parameters related to momenta of strings and satisfy equality $a +b +c = 1$, $g_\infty$ is a coupling constant, $|\cdot|_\infty$ denotes usual absolute value, $x$ is real variable related to the string world-sheet, and $\zeta$
is the Riemann zeta function.

Scattering amplitude between two $p$-adic strings was introduced as $p$-adic analog of  integral \eqref{3.1a},  i.e.
\begin{align}
 A_p (a, b) &= g_p^2 \,\int_{\mathbb{Q}_p}
|x|_p^{a-1}\, |1 -x|_p^{b-1}\, d_p x  \label{3.2a} \\  &= g_p^2 \, \frac{1 - p^{a - 1}}{1 -
p^{-a}}\, \frac{1 - p^{b - 1}}{1 - p^{-b}}\, \frac{1 - p^{c -
1}}{1 - p^{-c}}  ,    \label{3.2b}
\end{align}
where now $x$ is world-sheet variable described by $p$-adic (instead of real) numbers, while parameters $a, b, c $ remain their properties as in the case of ordinary strings.  According  to this definition of string amplitudes, it follows that $p$-adic and ordinary strings differ in description of their  world-sheet, i.e. by $p$-adic and real numbers, respectively.
Final expressions for amplitudes of ordinary \eqref{3.1b} and $p$-adic \eqref{3.2b} strings differ, but they are  connected by the Freund-Witten product formula
\begin{align}   A (a, b) =
 A_\infty (a, b) \prod_p A_p (a, b) = g_\infty^2 \,\prod_p g_p^2 = const.   \label{3.3}
\end{align}
In \eqref{3.3} is used the Euler product formula for the definition of the Riemann zeta function
\begin{align}
\zeta(s) = \prod_p \frac{1}{1- p^{-s}} , \quad s = \sigma + i \tau , \quad \sigma > 1 . \label{3.4}
\end{align}
Importance of product \eqref{3.3} consists in possibility to express complex ordinary string amplitude as product
of all inverse $p$-adic amplitudes which are simpler than the ordinary one, i.e. $ A_\infty (a, b) = const. \prod_p A_p^{-1} (a, b)$. It also  gives rise to think that not only ordinary strings may exist but also $p$-adic ones, or a new string (zeta string) that encompasses all $p$-adic effects.

One of the main achievements in $p$-adic string theory
is finding of an effective field description of
$p$-adic strings without explicit use of $p$-adic numbers. The corresponding {\em Lagrangian} is very simple
and exact. It describes not only four-point scattering amplitude
but also all higher ones at the tree-level.

 This Lagrangian for effective scalar field
$\varphi$, which describes open $p$-adic string (tachyon), is
\begin{align}
 {\cal L}_p (\varphi) = \frac{m_p^D}{g_p^2}\, \frac{p^2}{p-1} \Big[
-\frac{1}{2}\, \varphi \, p^{-\frac{\Box}{2 m_p^2}} \, \varphi +
\frac{1}{p+1}\, \varphi^{p+1} \Big] ,    \label{3.5}
\end{align}
where $p$
 is any prime number, $\Box = - \partial_t^2  + \nabla^2$ is the
$D$-dimensional d'Alembert operator, $m_p$ is string mass and  metric has signature $(- \, +
\, ...\, +)$. Kinetic term in \eqref{3.5} contains nonlocal operator with infinite number of
space-time derivatives
\begin{align}
  p^{-\frac{\Box}{2 m_p}} = \exp{\Big( - \frac{\ln{p}}{2
m_p}\, \Box \Big)} = \sum_{k \geq 0} \, \Big(-\frac{\ln p}{2 m_p}
\Big)^k \, \frac{1}{k !}\, \Box^k \,.  \label{3.6}
\end{align}
Potential $V_p(\varphi)$ is
\begin{align}
 V_p(\varphi) = - {\cal L}_p (\Box = 0) = \frac{m_p^D}{g_p^2}\, \frac{p^2}{p-1} \Big[
\frac{1}{2}\, \varphi^2  - \frac{1}{p+1}\, \varphi^{p+1} \Big]    \label{3.7}
\end{align}
and contains nonlinearity of the degree $p +1$.
%Note that there is some similarity of the Higgs  potential
%with this one when $p=3$.

The equation of motion for the scalar field $\varphi$ in Lagrangian \eqref{3.5} is
\begin{align}
 p^{-\frac{\Box}{2 m_p^2}}\, \varphi = \varphi^p ,    \label{3.8}
\end{align}
which  has two trivial
solutions $\varphi = 0$ and $\varphi =1$. There are also
inhomogeneous solutions resembling solitons, and for any spatial coordinate $x^i$ one has
\begin{align}
 \varphi (x^i) = p^{\frac{1}{2(p-1)}}\, \exp \Big( -
\frac{p-1}{2 m_p^2 \, p \ln p} (x^i)^2 \Big).        \label{3.9}
\end{align}

It is worth mentioning that taking  limit $p = 1 + \varepsilon \to 1$ in Lagrangian \eqref{3.5} one obtains
\begin{align}
\mathcal{L}_1 =\frac{m^D}{g^2} \Big[\frac{1}{2}\ \varphi \frac{\Box}{m^2} \varphi  + \frac{\varphi^2}{2}\
(\log{\varphi^2} - 1) \Big],    \label{3.5a}
\end{align}
and this new Lagrangian \eqref{3.5a} is related to ordinary scalar string, see \cite{gerasimov}.

Starting from Lagrangian \eqref{3.5}, many properties of $p$-adic string were investigated, e.g. see some references in
\cite{BD3} for more information.

\section{Zeta Strings}

Recall that starting from Lagrangian \eqref{3.5} and using methods of quantum field theory was obtained $p$-adic string scattering amplitude
\begin{align}
 A_p (a, b) = g_p^2 \, \frac{1 - p^{a - 1}}{1 -
p^{-a}}\, \frac{1 - p^{b - 1}}{1 - p^{-b}}\, \frac{1 - p^{c -
1}}{1 - p^{-c}} , \, \quad (a + b + c = 1)    \label{4.1}
\end{align}
which is the same as that derived  from integral expression \eqref{3.2a} with $p$-adic world sheet, see \cite{freund} for a review.

Since the scattering amplitude for the whole $p$-adic sector was obtained in the form
\begin{align}
A(a, b) = \prod_p A_p (a, b) =  g^2 \,  \frac{\zeta (a)}{\zeta (1- a)}\, \frac{\zeta (b)}{\zeta
(1 - b)}\, \frac{\zeta (c)}{\zeta (1 - c)} ,    \label{4.2}
\end{align}
 then the question arises about possibility to construct Lagrangian that might produce amplitude \eqref{4.2}.
If such Lagrangian exists, then to obtain it, we should start with Lagrangian for $p$-adic string \eqref{3.5} and then perform suitable summation or multiplication over all primes $p$ in such way to have a new Lagrangian with the Riemann zeta function.
Thus,  there are two approaches: additive and multiplicative.

 \subsection{Additive approach}

To use additive approach, note that the Riemann zeta function can be introduced in the following  ways:
\begin{align}
 &\zeta (s) = \sum_{n= 1}^{+\infty} \frac{1}{n^{s}} \,, \quad s = \sigma + i \tau \,, \,\,\, \sigma >1 , \label{4.3a}  \\
 &\frac{1}{\zeta (s)} = \sum_{n= 1}^{+\infty} \frac{\mu (n)}{n^{s}} \,, \quad s = \sigma + i \tau \,, \,\,\, \sigma >1
 \,, \,\,\, \mu (n) = \text{M\"obius function}  \label{4.3b}   \\
 &(1 - 2^{1-s})\zeta (s) = \sum_{n= 1}^{+\infty} (-1)^{n-1} \frac{1}{n^{s}} \,, \quad s = \sigma + i \tau \,, \,\,\, \sigma > 0. \label{4.3c}
\end{align}
It is well known that the above defined  Riemann zeta function has analytic continuation to the whole complex $s$-plane except the point $s = 1$, where it has a simple pole with residue 1.

 Note also that there is a sense to replace prime number $p$ by any natural number $n \geq 2$ in Lagrangian \eqref{3.5}, i.e. one can introduce
 \begin{align}
 {\cal L}_n (\varphi) = \frac{m_n^D}{g_n^2}\, \frac{n^2}{n-1} \Big[
-\frac{1}{2}\, \varphi \, n^{-\frac{\Box}{2 m_n^2}} \, \varphi +
\frac{1}{n+1}\, \varphi^{n+1} \Big] ,    \label{4.4}
\end{align}

Now we want to introduce a new  Lagrangian with a new field $\phi$ by the following sum:
\begin{align}
 L (\phi) &= \sum_{n = 1}^{+\infty} C_n\, {\cal L}_n  \nonumber \\
 &= \sum_{n = 1}^{+\infty} C_n\,
\frac{m_n^D}{g_n^2}\, \frac{n^2}{n-1} \Big[
-\frac{1}{2}\, \varphi \, n^{-\frac{\Box}{2 m_n^2}} \, \varphi +
\frac{1}{n+1}\, \varphi^{n+1} \Big] ,    \label{4.5}
\end{align}
whose concrete realization depends on the particular choice of  the coefficients $C_n$, masses $m_n$
and coupling constants $g_n$. To avoid a divergence problem of $1/(n-1)$ when  $n =1$, we  take that
$C_n$ is proportional to $n-1$. We also assume that $m_n$ and $g_n$ do not depend on $n$, and denote $m_n = m$
and $g_n = g$. Taking $C_n = \frac{n-1}{n^2} D_n$ we can rewrite \eqref{4.5} as
\begin{align}
 L (\phi) =  \frac{m^D}{g^2} \sum_{n = 1}^{+\infty} D_n\, \Big[
-\frac{1}{2}\, \phi \, n^{-\frac{\Box}{2 m^2}} \, \phi +
\frac{1}{n+1}\, \phi^{n+1} \Big].     \label{4.6}
\end{align}
One can easily see that term with $n=1$ is equal to zero and does not contribute to the sum in \eqref{4.6}, but its presence is useful to perform procedure required by definition of the Riemann zeta function.

In \cite{BD3}--\cite{BD9}, we introduced   new Lagrangians for the following values of coefficient $D_n$:  $D_n = 1 , \, \, D_n = n+1 , \, \,  D_n = \mu (n) , \, \, D_n = - \mu (n)\ (n+1) , \, \, D_n = (-1)^{n-1} , \, \, D_n = (-1)^{n-1}\ (n+1)$.

\subsubsection{Case $D_n =1$, \cite{BD3,BD4}.}

In this case we have
\begin{align}
 L (\phi) =  \frac{m^D}{g^2} \sum_{n = 1}^{+\infty}  \Big[
-\frac{1}{2}\, \phi \, n^{-\frac{\Box}{2 m^2}} \, \phi +
\frac{1}{n+1}\, \phi^{n+1} \Big].     \label{4.7}
\end{align}
Performing summation with application of the Euler formula \eqref{4.3a} and taking analytic continuation,
we can rewrite  \eqref{4.7} in the form
\begin{align}
 L (\phi) = - \frac{m^D}{g^2}  \Big[
\frac{1}{2}\, \phi \, \zeta\big(\frac{\Box}{2 m^2}\big) \, \phi + \phi + \frac{1}{2} \log{(1 -\phi)^2}\Big]     \label{4.8}
\end{align}
The potential $V(\phi) = - L (\Box = 0)$ is
\begin{align}
 V (\phi) =  \frac{m^D}{g^2}  \Big[
\frac{\zeta(0)}{2} \ \phi^2 + \phi + \frac{1}{2} \log{(1 -\phi)^2}\Big] ,     \label{4.9}
\end{align}
where $\zeta(0) = -1/2$. Potential  \eqref{4.9}
 has two  local maximums, that are unstable points: $V(0) = 0$ and $V (3) \approx 1.443 \frac{m^D}{g^2}$. It has  the following singularities:
 $\lim_{\phi \to 1} V (\phi) = -
\infty\,, \,\, \lim_{\phi \to \pm \infty} V (\phi) = - \infty $.

The corresponding equation of motion is
\begin{align}
 \zeta\Big( \frac{\Box}{2\, m^2}\Big) \phi =
\frac{\phi}{1 - \phi}   \label{4.10}
\end{align}
with  two trivial solutions:
$\phi =0$ and $\phi =3$. The
solution $\phi = 3$ follows from the Taylor expansion of the Riemann
zeta function operator
\begin{align}
\zeta\Big( \frac{\Box}{2\, m^2}\Big) = \zeta (0)  + \sum_{n
\geq 1} \frac{\zeta^{(n)} (0)}{n!}\, \Big( \frac{\Box}{2\,
m^2}\Big)^n . \label{4.11}
\end{align}

In the weak-field approximation $|\phi (x)| \ll 1$, equation of motion \eqref{4.10} becomes
\begin{align}
 \zeta\Big( \frac{\Box}{2\, m^2}\Big) \phi = \phi .   \label{4.12}
\end{align}

$\zeta\Big({\frac{\Box}{2 m^2}}\Big)$ can be regarded as a
pseudodifferential operator
\begin{align}
 \zeta\Big({\frac{\Box}{2 m^2}}\Big)\, \phi (x) =
\frac{1}{(2\pi)^D}\, \int_{\mathbb{R}^D} e^{ i x k}\,
\zeta\Big(-\frac{k^2}{2 m^2}\Big)\, \tilde{\phi}(k)\,dk   \label{4.12a}
\end{align}
with singularity at point $k^2 = - 2 m^2$.
Now mass spectrum is $M^2 = \mu m^2$, where $M^2 = - k^2 = k_0^2 -\overrightarrow{k}^2$, is determined by formula
\begin{align}
 \zeta\Big( \frac{M^2}{2\, m^2}\Big)  = 1   \label{4.12b}
\end{align}
and gives  many tachyon masses ($M^2 < 0$).

Note that one can  replace above $D_n = 1$ by $D_n = -1$. Then Lagrangian \eqref{4.8} and potential \eqref{4.9} will change their sign, while equation of motion \eqref{4.10} and mass spectrum formula \eqref{4.12b} will remain unchanged. This case may be more interesting than the previous one and will be elaborated elsewhere.

In \cite{BD4}, a more general Lagrangian was considered, i.e.
\begin{align}
L_h (\phi) = \frac{m^D}{g^2} \Big[-\frac{1}{2} \phi \zeta\Big(\frac{\Box}{2 m^2}+ h \Big) \phi + \mathcal{AC} \sum_{n=1}^\infty \frac{n^{-h}}{n+1} \phi^{n+1}     \Big] ,     \label{4.12c}
\end{align}
where $h$ is a real parameter (for $h=0$ Lagrangian \eqref{4.12c} reduces to \eqref{4.8}). The related equation of motion is
\begin{align}
\zeta\Big(\frac{\Box}{2m^2} + h\Big)\phi = \mathcal{AC} \sum_{n=1}^\infty n^{-h} \phi^n .  \label{4.12d}
\end{align}
Solution of equation of motion \eqref{4.12d} is investigated in \cite{prado}, where LHS is simplified by  an entire function of exponential type.

\subsubsection{Case $D_n =n+1$, \cite{BD5,BD6}.}

In this case, Lagrangian \eqref{4.6} becomes
\begin{align}
 L (\phi) =   \frac{m^D}{g^2} \Big[ - \frac{1}{2}\, \phi \, \sum_{n=
1}^{+\infty} \Big( n^{-\frac{\Box}{2 m^2} + 1} \, + \,
n^{-\frac{\Box}{2 m^2}} \Big) \, \phi  + \sum_{n= 1}^{+\infty}
\phi^{n+1} \Big] \label{4.13}
\end{align}
 and according to \eqref{4.3a} and analytic continuation we have
 \begin{align}
  L (\phi) = \frac{m^D}{g^2} \Big[ \, - \frac{1}{2}\,
 \phi \,  \Big\{ \zeta\Big({\frac{\Box}{2\, m^2}  -
 1}\Big)\, + \, \zeta\Big({\frac{\Box}{2\, m^2} }\Big) \Big\} \, \phi \,  + \,   \frac{\phi^2}{1 - \phi} \,
 \Big]\,. \label{4.14}
 \end{align}

Since $\zeta (-1) = - 1/12$ and $\zeta (0) = - 1/2$, the corresponding potential is
\begin{align}
  V (\phi) = -L(\Box = 0) =  \frac{m^D}{g^2}  \   \frac{7 \phi - 31}{24\ (1 - \phi)}
 \ \phi^2
 \,, \label{4.15}
 \end{align}
 with properties: $V (0) = V (31/7) = 0\,, \,\, V (1\pm 0) = \pm \infty\,, \,\, V (\pm \infty) = -
 \infty$. At $\phi = 0$ potential has local maximum.

The equation of motion is
\begin{align}
 \Big[ \zeta\Big({\frac{\Box}{2\, m^2}  -
 1}\Big)\, + \, \zeta\Big({\frac{\Box}{2\, m^2} }\Big) \Big] \, \phi
 = \frac{\phi ((\phi - 1)^2 + 1)}{(\phi - 1)^2}\,, \label{4.16}
\end{align}
which has only  $\phi = 0$ as a real constant  solution.

The weak field approximation of \eqref{4.16} is
\begin{align}
 \Big[ \zeta\Big({\frac{\Box}{2\, m^2}  -
 1}\Big)\, + \, \zeta\Big({\frac{\Box}{2\, m^2} }\Big) - 2 \Big] \, \phi
 = 0 , \label{4.17}
\end{align}
which implies condition on the mass spectrum
\begin{align}
  \zeta\Big(\frac{M^2}{2\, m^2}  -
 1\Big)\, + \, \zeta\Big({\frac{M^2}{2\, m^2} }\Big)
 = 2\,. \label{4.18}
 \end{align}
From (\ref{4.18})  follows that there are no finite solutions  $M^2 > 2 m^2$  and there are many tachyon solutions  $M^2 <  0$.

\subsubsection{Case $D_n = \mu(n)$, \cite{BD7}.}

Related Lagrangian is
\begin{align}
 L_\mu (\phi) =  \frac{m^D}{g^2} \sum_{n = 1}^{+\infty}  \Big[
-\frac{1}{2}\, \phi \, \mu(n)\ n^{-\frac{\Box}{2 m^2}} \, \phi +
\frac{\mu(n)}{n+1}\, \phi^{n+1} \Big].     \label{4.19}
\end{align}
where  $ \mu (n)$ is the M\"obius function:
\begin{equation}
\mu (n)= \left \{ \begin{array}{lll} 0 , \quad &  n = p^2 m \\
(-1)^k , \quad & n = p_1 p_2 \cdots p_k ,\,\,  p_i \neq p_j  \\
1 , \quad & n = 1, \,\,  (k=0)
\end{array} \right.   \label{4.20}
\end{equation}

Taking into account zeta
function  by expression \eqref{4.3b}
 one can rewrite  Lagrangian \eqref{4.19} as
\begin{align}
 L_\mu (\phi) =  \frac{m^D}{g^2}   \Big[
-\frac{1}{2}\, \phi \, \frac{1}{\zeta\big(\frac{\Box}{2 m^2}\big)} \, \phi + \mathcal{AC}
\int_{0}^\phi \mathcal{M}(\phi)\ d\phi \Big],     \label{4.21}
\end{align}
where $\mathcal{AC}$ denotes analytic continuation and $\mathcal{M}(\phi) = \sum_{n=1}^{+\infty} \mu(n) \phi^n = \phi - \phi^2 - \phi^3 - \phi^5 + \phi^6 -\phi^7 + \phi^{10} - \ldots$

The corresponding potential, equation of motion and formula for mass spectrum are, respectively:
\begin{align}
&V_\mu (\phi) = - \frac{m^D}{g^2} \Big[\phi^2  + \mathcal{AC} \int_{0}^\phi \mathcal{M}(\phi)\ d\phi  \Big] , \label{4.22a} \\
&\frac{1}{\zeta\big(\frac{\Box}{2 m^2}\big)}\ \phi = \mathcal{M}(\phi) , \quad \zeta\big(\frac{M^2}{2 m^2}\big) = 1. \label{4.22b}
\end{align}
In mass spectrum, there are only tachyons.

\subsubsection{Case $D_n = - \mu(n)\ (n+1)$, \cite{BD8}.}

Related Lagrangian is
\begin{align}
 L_{-\mu} (\phi) =  \frac{m^D}{g^2}\  \sum_{n=1}^{+\infty} \phi \Big[ \frac{\mu (n)}{2} \
n^{-\frac{\Box}{2 m^2 } + 1}  +  \frac{\mu (n)}{2} \ n^{-\frac{\Box}{2 m^2}} -  \mu (n) \, \phi^{n-1} \Big] \phi ,  \label{4.23}
\end{align}
 where  $ \mu (n)$ is the M\"obius function. Using procedure as in the previous case, one can rewrite Lagrangian \eqref{4.23} in the form
\begin{align}
  L_{-\mu} (\phi) = \frac{m^D}{g^2} \Big\{
\frac{1}{2}\, \phi\, \Big[ 1/\zeta \big(\frac{\Box}{2 m^2 } - 1\big) +
1/\zeta \big(\frac{\Box}{2m^2 }\big) \Big] \, \phi -  \phi^{2} \mathcal{AC}\ F (\phi)
\Big\}  \label{4.24}
\end{align}
 where $F (\phi) =  \sum_{n=1}^{+\infty} \mu (n) \ \phi^{n-1}  =  (1 - \phi - \phi^2 -\phi^4 + \phi^5 - \ldots ) .$ Infinite sum  $\sum_{n=1}^{+\infty} \mu (n) \ \phi^{n-1}$ is convergent when $|\phi| < 1 .$

 Taking into account that $\zeta(-1) = -1/12$  and $\zeta(0) = -1/2$, the potential is
\begin{align}
V_{-\mu} = \frac{m^D}{g^2} \big[7 + \mathcal{AC} \ F (\phi) \big] \phi^2 . \label{4.25}
\end{align}
The related equation of motion and mass spectrum formula are:
\begin{align}
  \Big[ \zeta^{-1} \big(\frac{\Box}{2 m^2 } - 1\big) &+
\zeta^{-1} \big(\frac{\Box}{2m^2 }\big) \Big] \, \phi = 2 \phi  \mathcal{AC}\ F (\phi) + \phi^{2} \mathcal{AC}\ F' (\phi) , \label{4.2a} \\
 &\zeta^{-1} \big(\frac{M^2}{2 m^2 } - 1\big) +
\zeta^{-1} \big(\frac{M^2}{2m^2 }\big)  = 2 .  \label{4.25b}
\end{align}

\subsubsection{Case $D_n = (-1)^{n-1}$.}

A new example is based on \eqref{4.3c}
\begin{align}
 \sum_{n= 1}^{+\infty} (-1)^{n-1} \frac{1}{n^{s}} =  (1 - 2^{1-s})
\, \zeta (s), \quad s = \sigma + i \tau \,, \quad \sigma
> 0  \label{4.25}
\end{align}
which has analytic continuation to the entire complex $s$ plane with the corresponding analytic  expansion \cite{wikipedia}
\begin{equation}
(1 - 2^{1-s}) \, \zeta (s) = \sum_{n=0}^\infty \frac{1}{2^{n+1}}
\, \sum_{k=0}^n (-1)^k \,
\binom{n}{k} (k +1)^{-s} \,. \label{4.26}
\end{equation}
 At point $s = 1$, one obtains
\begin{align}
\lim_{s\to 1} (1 - 2^{1-s}) \ \zeta
(s)\ = \ \sum_{n= 1}^{+\infty} (-1)^{n-1} \frac{1}{n} \ = \ \log 2.   \label{4.27}
\end{align}

Applying formula \eqref{4.25} to Lagrangian \eqref{4.6} and using analytic continuation, we have
\begin{align}
L = \frac{m^D}{g^2} \ \Big[ - \frac{1}{2} \phi \Big( 1 - 2^{1- \frac{\Box}{2
m^2}} \Big) \zeta\Big(\frac{\Box}{2 m^2}\Big) \phi \ + \phi -
\frac{1}{2} \ \log(1 + \phi)^2 \Big] .  \label{4.28}
\end{align}

The potential is
\begin{align}
 V(\phi) = - L (\Box = 0) = \frac{m^D}{g^2} \Big[ \frac{1}{4} \, \phi^2 - \phi
+ \frac{1}{2} \log (1 +\phi)^2 \Big],  \label{4.29}
\end{align}
which has one local maximum $V(0) = 0$ and one local minimum $V(1) \approx - 0.057 \frac{m^D}{g^2}$.
It is singular at $\phi = -1$, i.e. $V(-1) = -
\infty$, and $V (\pm \infty) = + \infty$.  The equation of motion
is
\begin{align}
 \Big( 1 - 2^{1- \frac{\Box}{2 m^2}} \Big) \zeta\Big(\frac{\Box}{2
m^2}\Big) \phi = \frac{\phi}{1+ \phi},   \label{4.30}
\end{align}
which has two constant solutions: $\phi = 0$ and $\phi = 1$. Formula for the mass spectrum is
\begin{align}
 \Big( 1 - 2^{1- \frac{M^2}{2 m^2}} \Big) \zeta\Big(\frac{M^2}{2
m^2}\Big)  = 1.    \label{4.31}
\end{align}

\subsubsection{Case $D_n = (-1)^{n-1} (n +1)$, \cite{BD9,BD8}.}

Applying $D_n = (-1)^{n-1} (n +1) $ and formula \eqref{4.3c} to Lagrangian (\ref{4.6}), and using analytic
continuation we obtain
 \begin{align} \nonumber L = & - \frac{m^D}{g^2} \Big\{ \,  \frac{1}{2}\,
 \phi \,  \Big[ \, \Big(1 - 2^{2 - \frac{\Box}{2 m^2}}\Big)\, \zeta\Big({\frac{\Box}{2\, m^2}  -
 1}\Big)\, \\ & + \,  \Big(1 - 2^{1 - \frac{\Box}{2 m^2}}\Big)\, \zeta\Big({\frac{\Box}{2\, m^2} }\Big)
 \Big] \, \phi \,  - \,   \frac{\phi^2}{1 + \phi} \,
 \Big\}\,. \label{4.32} \end{align}
 The  related potential, equation of motion and mass spectrum formula are, respectively:
 \begin{align}
 &V(\phi)  = \frac{m^D}{g^2} \Big(\frac{3}{8}  - \frac{1}{1 + \phi} \Big) \phi^2 \ , \label{4.33} \\
 &\Big[ \Big(1 - 2^{2 - \frac{\Box}{2 m^2}}\Big)\ \zeta\Big({\frac{\Box}{2\ m^2}  -
 1}\Big) +   \Big(1 - 2^{1 - \frac{\Box}{2 m^2}}\Big)\ \zeta\Big(\frac{\Box}{2\, m^2} \Big)   \Big] \phi \nonumber \\
 &= \frac{2\phi + \phi^2}{(1+\phi)^2} ,  \label{4.34} \\
 &\Big(1 - 2^{2 - \frac{M^2}{2 m^2}}\Big)\ \zeta\Big({\frac{M^2}{2\ m^2}  -
 1}\Big) +   \Big(1 - 2^{1 - \frac{M^2}{2 m^2}}\Big)\ \zeta\Big(\frac{M^2}{2\, m^2} \Big) = 2 .  \label{4.35}
 \end{align}

\subsection{Multiplicative approach}

 Let us note that Lagrangian for $p$-adic strings \eqref{3.5} can be rewritten as follows (see \cite{BD7}):
 \begin{align}
 \mathcal{L}_p (\varphi) &= \frac{m_p^D}{g_p^2}\, \frac{p^2}{p-1} \Big[
-\frac{1}{2}\, \varphi \, p^{-\frac{\Box}{2 m_p^2}} \, \varphi +
\frac{1}{p+1}\, \varphi^{p+1} \Big]  \label{4.2.1} \\
&= \frac{m^D}{g^2}\, \frac{p^2}{p^2-1} \Big\{
\frac{1}{2}\, \varphi \, \Big[ \Big(1 - p^{-\frac{\Box}{2 m^2}+1}
\Big)  + \Big( 1 - p^{-\frac{\Box}{2 m^2}}\Big) \Big]\, \varphi \nonumber \\
 &- \varphi^2 \Big(1 - \varphi^{p-1} \Big) \Big\} , \label{4.2.2}
\end{align}
where we take mass $m_p = m$   and coupling constant $g_p = g .$
Now we introduce a new Lagrangian $\mathcal{L} (\phi)$ by the following steps:
\begin{align}
\mathcal{L}_p (\varphi) \to \prod_p \mathcal{L}_p (\varphi) \to \mathcal{AC} \prod_p \mathcal{L}_p (\varphi) =  \mathcal{L} (\phi),    \label{4.2.3}
\end{align}
where $\mathcal{AC}$  means  analytic continuation, and introduced new scalar field is denoted by $\phi$. The explicit form of
$\mathcal{L} (\phi)$ is
\begin{align}
  {\mathcal L} = \frac{m^D}{g^2}\, \zeta (2)\, \Big\{ \frac{1}{2} \,
\phi \Big[ \frac{1}{\zeta\Big( \frac{\Box}{2 m^2} - 1 \Big)} +
\frac{1}{\zeta\Big( \frac{\Box}{2 m^2} \Big)}\Big] \, \phi - \phi^2
  \mathcal{AC}\ G(\phi) \Big\} , \label{4.2.4}
\end{align}
where $G (\phi) =  \prod_p (1 -\phi^{p-1})$.
 Infinite product
$\prod_p (1 -\phi^{p-1})$ is convergent if $|\phi| <
1 $. One can easily see that $G (0) = 1$ and $G (1) = G (-1) = 0$.

The corresponding equation of motion  is
\begin{align}
 \Big[1/ \zeta\Big( \frac{\Box}{2 m^2 } - 1 \Big) + 1/\zeta
\Big( \frac{\Box}{2 m^2} \Big)\Big] \, \phi  = 2 \phi \mathcal{AC}\ G (\phi)
+ \phi^2 \mathcal{AC}\ G' (\phi)  \label{4.2.5}
\end{align}
and has $\phi = 0$ as a trivial solution. In the weak-field
approximation (i.e. $|\phi (x)| \ll 1$), equation of motion \eqref{4.2.5}  becomes
\begin{align}
 \Big[ 1/\zeta \Big( \frac{\Box}{2 m^2} - 1 \Big) + 1/\zeta
\Big( \frac{\Box}{2 m^2} \Big)\Big] \, \phi  = 2 \phi .
\end{align}

The potential   $ V (\phi) = - {\mathcal L} ( \Box = 0)$ is
\begin{align}
  V (\phi) = \frac{m^D}{g^2}\ \zeta (2)\ [ 7 + \mathcal{AC}\ G(\phi)] \ \phi^2 ,
\end{align}
where  $\zeta (- 1) = - 1/12$ and $\zeta (0) = - 1/2$ are taken into account. This
potential has local minimum  $V (0) = 0$ and values $V (\pm 1) =
7\, m^D$. To explore behavior of $V (\phi)$ for all $\phi \in
\mathbb{R}$ one has first to investigate properties of the
function $ G(\phi)$.

Mass spectrum of $M^2$ is determined by solutions of equation
\begin{align}
\zeta^{-1} \big(\frac{M^2}{2 m^2 } - 1\big) +
\zeta^{-1} \big(\frac{M^2}{2m^2 }\big)  = 2 .
\end{align}
 There are  many tachyon solutions.

{\bf Remark.} \  The difference between Lagrangians ${\mathcal L}$ \eqref{4.2.4} (multiplicative
approach) and $L_{-\mu}$ \eqref{4.24} (additive approach) is practically in functions $G
(\phi)$ and $F (\phi)$. Since
\begin{align}
 G (\phi) = \prod_p \big(1 - \phi^{p-1} \big) = 1 - \phi
-\phi^2 + \phi^3 - \phi^4 + ...
\end{align}
 and
\begin{align}
 F (\phi) = \sum_{n =1}^\infty \mu (n) \phi^{n-1} =  1 - \phi - \phi^2 - \phi^4 + \phi^5 - \ldots
\end{align}
 it follows that these functions approximatively have the same
behavior for $|\phi| \ll 1$. Hence, in  the weak-field approximation
these Lagrangians  describe the same scalar field.

\section{Concluding Remarks}

%\section{Conclusions}

In this review article, seven possible Lagrangians for zeta strings are presented. These Lagrangians contain non-polynomial nonlinearity and space-time nonlocality. This nonlocality is encoded in the Riemann zeta function with d'Alembert operator $\Box$ as its argument. The corresponding potentials, equations of motion and the mass spectrum formulas are also presented. All of these seven models of zeta strings contain the Riemann zeta function and have some intriguing properties that deserve further investigation.

Zeta-string model presented in Section 4.1.5 is a new one and  further investigation will be presented elsewhere. In \cite{BD3}, Lagrangian for coupled open and closed zeta string is constructed.

Instead of the Riemann zeta function one can try construction of Lagrangians with some similar functions. In \cite{prado}, embedding of the Dirichlet L-function was proposed.

 A sum of Lagrangian $\mathcal{L}_1$ \eqref{3.5a} with any of the above zeta-string Lagrangians should give some information
 about connection between ordinary and zeta  strings.

Note that an interesting approach towards a field theory and cosmology based on the Riemann zeta function and its generalizations is proposed in \cite{AV}, see also  \cite{BD1}. In \cite{BD2}, the vacuum state of the adelic harmonic oscillator is connected with the Riemann zeta function in the form of its functional relation.


\begin{thebibliography}{99}
\medskip
\begin{footnotesize} % Please do not modify

\bibitem{GSW}
M. B. Green, J. H. Schwarz and E. Witten,
{\it Superstring Theory} (Cambridge University Press, 1987).

\bibitem{freund} L. Brekke and P.G.O. Freund, ``$p$-Adic numbers in
physics'', {\em Phys. Rep.} {\bf 233} (1993) 1.

\bibitem{VVZ} V.S. Vladimirov, I.V. Volovich and E.I.
Zelenov, {\em $p$-Adic Analysis and Mathematical Physics} (World
                 Scientific, Singapore, 1994).

\bibitem{BD1} B. Dragovich, A.Yu. Khrennikov, S.V. Kozyrev, I.V. Volovich and E.I. Zelenov, ``$p$Adic mathematical physics: the first 30 years'', {\em $p$-Adic Numbers Ultrametric Anal. Appl.} {\bf 9} (2017)  87--121; arXiv:1705.04758 [math-ph].

\bibitem{BD2} B. Dragovich, ``Adelic harmonic oscillator'', {\em Int.J.Mod.Phys.} {\bf A10} (1995) 2349-2365;
arXiv:hep-th/0404160.

\bibitem{BD3} B. Dragovich, ``Zeta Strings'',
arXiv:hep-th/0703008 (2007).

\bibitem{BD4} B. Dragovich, ``Zeta-Nonlocal Scalar
Fields'',  {\em Theor. Math. Phys.} {\bf 157}
(2008) 1671--1677; arXiv:0804.4114v1 [hep-th].

\bibitem{BD5}  B. Dragovich,  ``Lagrangians with
Riemann Zeta Function'',   {\em Romanian J. Physics
} {\bf 53} (2008) 1105-1110; arXiv:0809.1601v1[hep-th].

\bibitem{BD6} B. Dragovich,  ``Some Lagrangians with Zeta Function Nonlocality'';
arXiv:0805.0403v1[hep-th],  (2008).

\bibitem{BD7} B. Dragovich,  ``Towards Effective
Lagrangians for Adelic Strings'',  {\em Fortschr.
Phys.} {\bf 57} (2009) 546--551;  arXiv:0902.0295v1 [hep-th].

\bibitem{BD8}  B. Dragovich,  ``Nonlocal Dynamics of
$p$-Adic Strings'',  {\em Theor. Math. Phys.} {\bf 164}
(2010) 1151--1155; 	arXiv:1011.0912 [hep-th].

\bibitem{BD9} B. Dragovich,  ``The $p$-Adic Sector of
Adelic Strings'',  {\em Theor. Math. Phys.} {\bf 163} (2010) 768--773; 	arXiv:0911.3625 [hep-th].


\bibitem{gelfand} I.M. Gelfand, M.I. Graev and I.I. Pyatetskii-Shapiro, {\em Representation Theory and Automorphic Functions}
(Saunders, Philadelphia, 1969).

\bibitem{gerasimov}  A.A. Gerasimov and S.L. Shatashvilli, ``On exact tachyon potential in open string field theory'', {\em JHEP} {\bf } (2000) 034;  arXiv:hep-th/0009103.


 \bibitem{prado}  A. Ch\'avez, H. Prado and E.G. Reyes, ``The Borel transform and linear nonlocal equations:
applications to zeta-nonlocal field models''; 	arXiv:1907.02617 [math-ph],  (2019).

 \bibitem{wikipedia} J. Sondow,  ``Analytic continuation of Riemann’s zeta function and values at negative integers via Euler’s transformation of series'', {\em Proc. Amer. Math. Soc.} \textbf{120}, 421--424 (1994).

\bibitem{AV} I.Ya. Aref’eva and I.V. Volovich, ``Quantization of the Riemann zeta-function and cosmology'', {\em Int. J. Geom. Meth. Mod. Phys.} {\bf 4} (2007) 881; arXiv:hep-th/0701284.

\end{footnotesize} % Please do not modify
\end{thebibliography}
\end{document}